\documentclass{jaa}
\usepackage{natbib}
\usepackage{graphicx}
\usepackage{url}
\usepackage{epstopdf}
\newcommand{\Msun}{\ensuremath{M_{\odot}\,}}

\newcommand{\flux}{erg\,cm$^{-2}$\,s$^{-1}$}
\newcommand{\lum}{erg\,s$^{-1}$}
\newcommand{\s}{$\sim$}

\newcommand{\pyr}{$yr^{-1}$}
\newcommand{\pms}{$\pm\,$}
\newcommand{\arcs}{$''$}
\newcommand{\arcm}{$'$}

\newcommand{\tim}{$\times\,$}

\newcommand{\bv}{\ensuremath{B\!-\!V}}

\newcommand{\A}{$\mathring{A}$}

\begin{document}\sloppy

\title{Imaging and photometric studies of NGC~1316 (Fornax~A) using \\Astrosat/UVIT}

\author{Nilkanth D. Vagshette\textsuperscript{1,*}, Sachindra Naik\textsuperscript{2}, Neeraj Kumari\textsuperscript{2} \and Madhav K. Patil\textsuperscript{3}}
\affilOne{\textsuperscript{1}Department of Physics and Electronics, Maharashtra Udayagiri Mahavidyalaya, Udgir 413 517, India\\}
\affilTwo{\textsuperscript{2}Astronomy and Astrophysics Division, Physical Research Laboratory, Navrangpura, Ahmedabad 380 009, India\\}
\affilThree{\textsuperscript{3}School of Physical Sciences, Swami Ramanand Teerth Marathwada University, Nanded 431 606, India}

\twocolumn[{

\maketitle

\corres{nilkanth1384@gmail.com}

\msinfo{28 October 2020}{13 January 2021}

\begin{abstract}
We present imaging and photometric studies of the radio galaxy NGC~1316 (Fornax~A) using high spatial resolution near-ultraviolet (NUV) and far-ultraviolet (FUV) imaging telescopes of the first Indian multi-wavelength space observatory {\it AstroSat}. The residual maps of UV emission obtained from the subtraction of smooth models witness peculiar features within the central few kpc (1-2 kpc) region. The spatial correspondence between the radio emission maps and FUV imaging study reveal that the UV emitting sources are displaced away from the centre by the AGN outburst (radio jet). The presence of rims and clumpy structures in the outskirt of this galaxy delineate that the galaxy has acquired a large fraction of gas through merger-like events and is still in the process of settling. The estimates of the star formation rates (SFR) using FUV and NUV luminosities are found to be 0.15 M$_\odot$yr$^{-1}$ and 0.36 M$_\odot$yr$^{-1}$, respectively, and provide the lower limit due to the screen effect. The estimated lower rates of SFR in this galaxy probably represent its quenching due to the AGN driven outflows emanating from the central engine of NGC~1316.

\end{abstract}


}]

\doinum{12.3456/s78910-011-012-3}
\artcitid{\#\#\#\#}
\volnum{000}
\year{0000}
\pgrange{1--}
\setcounter{page}{1}
\lp{1}

\section{Introduction}
NGC~1316 (Fornax~A) is a nearby (z=0.00587) giant peculiar S0 galaxy hosting numerous tidal tails, shells, unusual dust patches all embedded within a much larger outer envelop of stars and a prominent dust lane oriented along its optical minor axis \citep{1983ApJ...274..534M,2013RAA....13..885D}. In addition to the intricate dust patches and shells, NGC~1316 also hosts filamentary, nebular emission features, ripples, arcs and several complex filamentary loops of other phases of inter stellar medium (ISM) \citep{1988ApJ...328...88S}. The faint tidal tails -- wisps and shells of stars evident around the galaxy suggest that they have been torn from their original locations and flung into the intergalactic space through complex gravitational effects. \citet{1980ApJ...237..303S} reported the presence of a compact disk of gas near its center that has different orientation and much faster rotation relative to the stars. All of these signs of complex dust features, shells, loops and other sub-structures evident in NGC~1316 collectively point to its violent past built-up through the merger of several smaller dust rich galaxies in distant past \citep{2002MNRAS.330..547T}. Schweizer (1980) based on the expansion time of the outer stellar loops and \citet{2001MNRAS.328..237G} on the spread in the ages of globular clusters demonstrate that the mergers might have happened between ~1 to ~3 Gyr ago and then NGC 1316 has continued accretion of smaller satellite galaxies \citep{2017ApJ...839...21I}.

NGC~1316 is one of the nearest Central Dominant (cD) giant radio galaxy, at a luminosity distance of $\sim$25 Mpc \citep{2006PASP..118.1711W} that exhibit filamentary low-ionization nuclear emission, an unresolved bright nucleus in UV bands, and interestingly hosts a strong radio core with steep spectrum and dual jets. The double-lobe radio continuum source from this galaxy extend well beyond its telescopic field-of-view, spanning over several degrees on the sky \citep{1983A&A...127..361E, 2001MNRAS.326.1076D}, and exhibit sharp edges in its outer part \citep{1983A&A...127..361E}. The S-shaped nuclear radio jet in the central region of NGC~1316 appears to bend at south of NW shell and north of SE blob due to the interaction between the radio jet and the ISM \citep{2019PASJ...71...85M}. In the photographic plates. NGC~1316 appears to interact with a close companion spiral galaxy NGC~1317 on its North, however, this companion is not dominant enough to distort NGC~1316 at the observed scales. Though formation of NGC~1316 through merging episodes is unquestionable, the composition and characteristics of the multi-phase ISM in the galaxy is not yet fully understood. Relative to its very large stellar content $ \sim 6 \pm 2 \times 10^{11} M_\odot $, the dust content of NGC~1316 is very low $\sim 2 \times 10^6 M_\odot$, \citep{2007ApJ...663..866D,2013RAA....13..885D} and is probably due to a ~10:1 merger between a dominant, dust-poor early-type galaxy and a smaller, gas-rich spiral \citep{2010ApJ...721.1702L}.

Our earlier study \citep{2013RAA....13..885D} of wavelength-dependent nature of dust extinction over the range near-infrared (NIR) to UV revealed that the dust grains in NGC~1316 posses identical physical and chemical characteristics as that of the canonical grains in the Milky Way and was confirmed through the parallel extinction curve. However, the smaller values of R$_\lambda$ indicate that the dust grains in the environment of NGC~1316 are relatively smaller than the canonical grains \citep{2007A&A...461..103P}. This study also quantifies the dust mass assuming screening effect of dust  \citep{2007A&A...461..103P,2012NewA...17..524V}, to be \s 2.12 \tim 10$^5$ \Msun, while that estimated from the MIPS far-infrared flux was found to be 3.21 \tim 10$^7$ \Msun. 

As ultraviolet (UV) and Far-Infrared (FIR) emission originate from the young stars, therefore, the UV and IR luminosities are often used in literature for investigating star formation histories of external galaxies \citep{2010MNRAS.409L...1B,2012ARA&A..50..531K}. Particularly, the UV continuum at $\lambda$ $<$ 2000 \A \ provides with the more reliable estimates of star formation rates \citep{2007ApJS..173..267S} . As a result, several attempts have been made in the past to better understand the star formation in external galaxies using UV observations. However, the Earth atmosphere acts as a main hindrance in acquiring UV data on such galaxies, as a result, we need to rely on space-based observatories like {\it GALEX} and/or {\it AstroSat} for observing such galaxies in UV bands. Therefore, the UV observations of NGC~1316 acquired using {\it AstroSat} observatory are crucial for understanding the star formation and hence evolution of the galaxy.

The merger episodes in NGC~1316 fuel the central supermassive black hole and transform it into a strong radio source, an Active Galactic Nucleus \citep{2006MNRAS.365...11C,2013MNRAS.428..641O,2016MNRAS.456..300O,2016MNRAS.458.2371R,2016MNRAS.461.1885V,2017MNRAS.466.2054V,2019MNRAS.485.1981V,2012MNRAS.421..808P,2015Ap&SS.359...61S}. The mergers drive the external gas inflow into the central part of the galaxy and hence also activate the star formation in  central region \citep{2008ApJ...682L..13H}. But the ignited AGN develop powerful jets, which then heat and blow up the surrounding ISM and hence suppress the star formation \citep{2006MNRAS.365...11C,2012ARA&A..50..455F}. As a result, the detailed study of the star formation and interaction of radio jets with the surrounding environment in merger remnants is interesting. Therefore, NGC~1316 is one of the suitable candidates of recent mergers to investigate star formation and interplay between the AGN and ISM.

Using recent high resolution radio observations with MeerKAT, \citet{2019A&A...628A.122S} have demonstrated that NGC~1316 hosts a significant fraction of H~I emission distributed in a variety of scales i.e. all the way from its center till the large-scale environment. \citet{2019A&A...628A.122S} found that the H~I emission detected in the central $\sim$5 kpc as additional component to the complex interstellar medium in this region: including dust \citep{2010Ap&SS.327..267C,2013RAA....13..885D}, dust emission \citep{2010ApJ...721.1702L},  molecular gas \citep{2001A&A...376..837H,2019PASJ...71...85M}, ionised gas \citep{1980ApJ...237..303S,2019PASJ...71...85M}, and X-ray emitting gas (this paper). \citet{2019A&A...628A.122S} have also revealed the extension of H~I emission up to 70–150 kpc and oriented along both the tails of NGC~1316 with a total mass content of 7 \tim 10$^8$\Msun, roughly 14 times larger than the past detection. Interestingly, both these H~I tails of NGC~1316 show a spatial association with the complex optical tidal features confirming their common origin through the merging of a gas-rich progenitor system. They further propose that the merger was so significant that the tidal forces pulled a large fraction of gas and stars out to larger radii apparent in the form of optical tails. 

Further, the low frequency observation between 154 and 1510 MHz shows that both the radio lobes exhibits a steep spectrum \citep{2015MNRAS.446.3478M}. These steep spectrum lobes are basically characterised by optically thin synchrotron radio emission from jets and they are associated with very massive black holes in the central early type galaxies \citep{2017ApJ...842...95M}.

In this paper, we present a detailed study of NGC~1316 galaxy using the high spatial resolution {\it AstroSat} ultra-violet imaging telescope (UVIT) data. The paper is structured as follows: Section~2 reports the observation and data analysis methods, whereas Section~3 describes the results obtained from the UV imaging analysis. Section~4 discusses the star-formation and AGN outflow in NGC~1316 and lastly, Section~5 summarises the results obtained from the present study.

\begin{figure*}
\centering\includegraphics[height=.33\textheight]{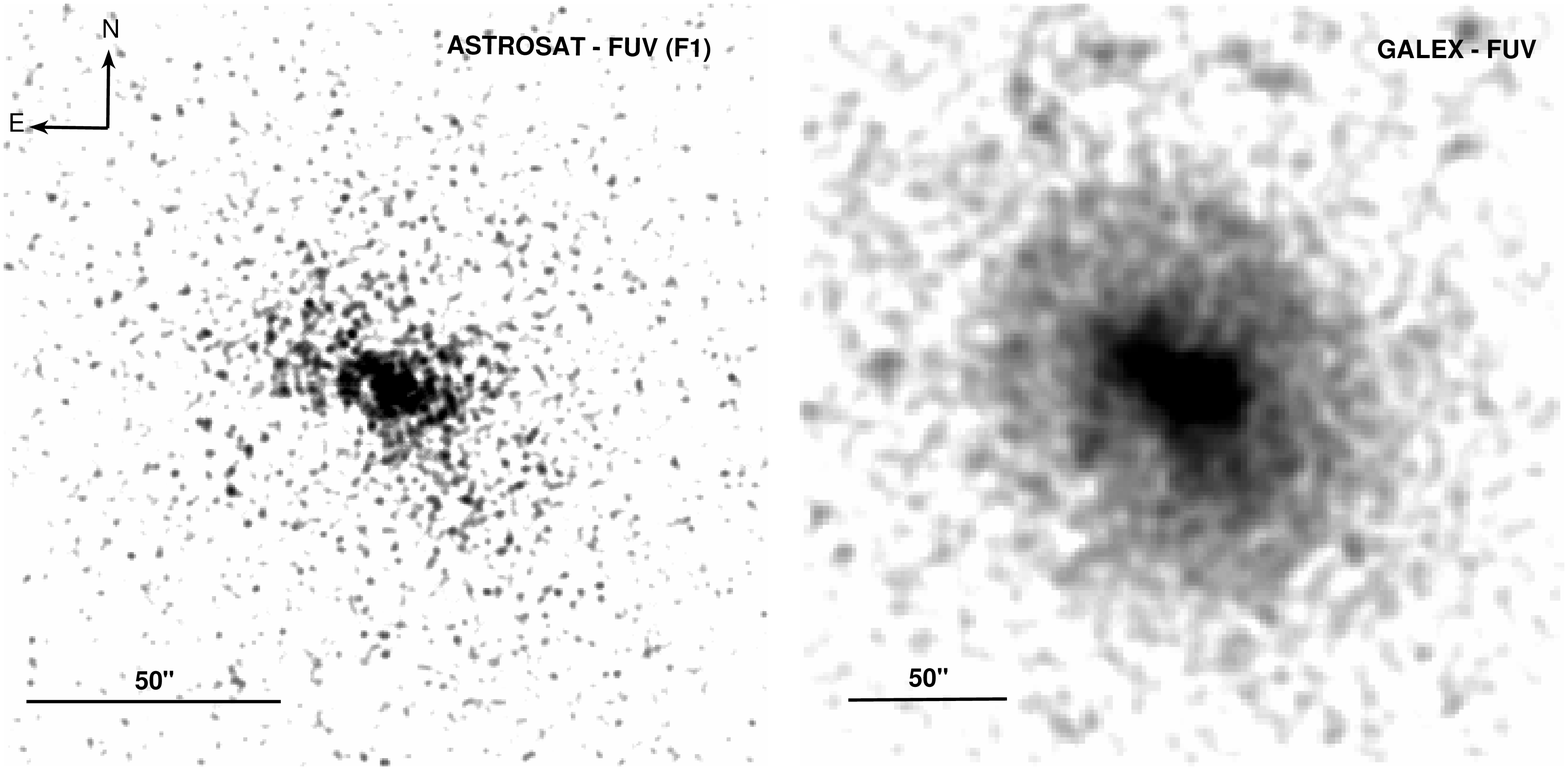}
\centering\includegraphics[height=.33\textheight]{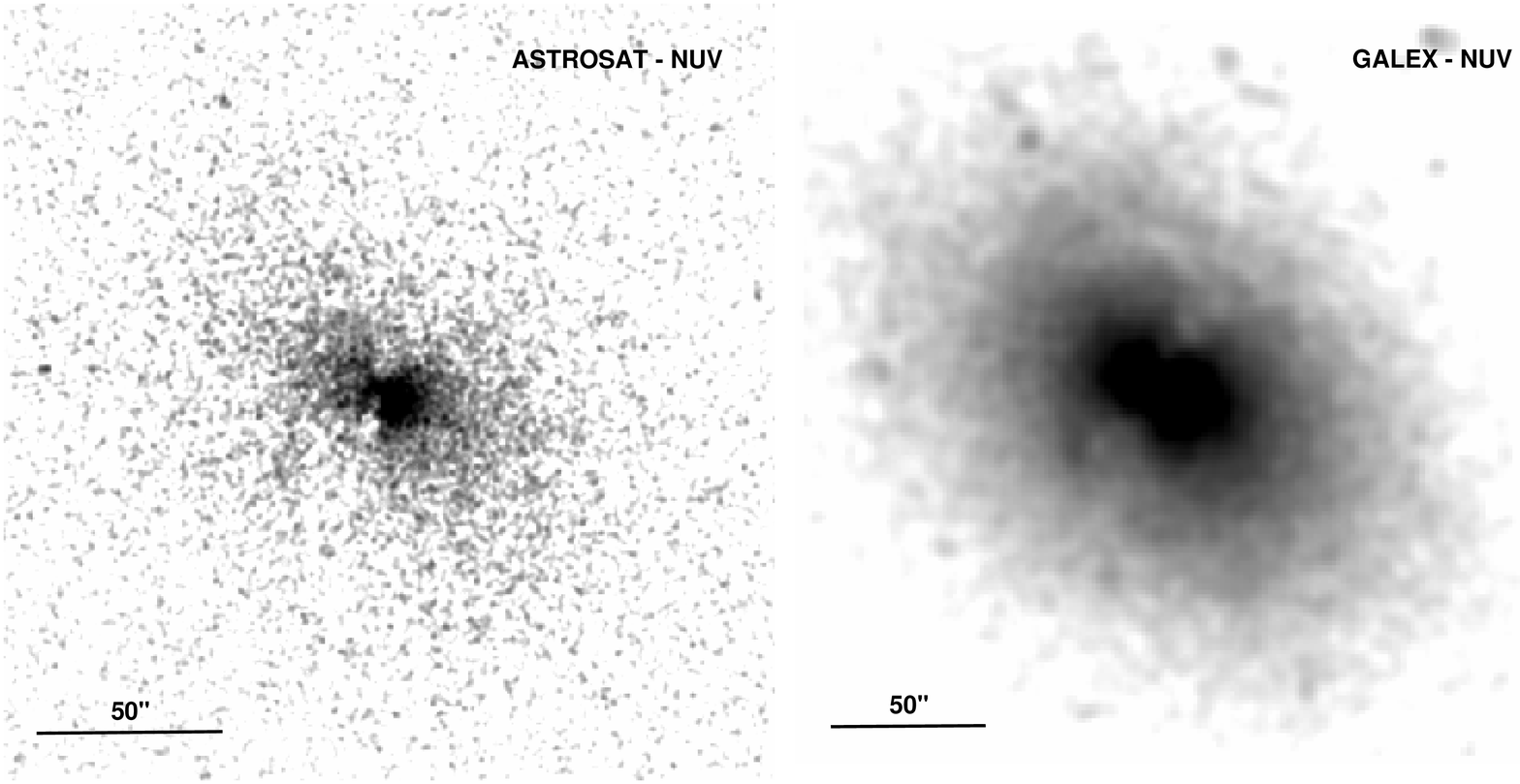}\\
\caption{The upper panel represents the FUV images of NGC~1316 using ASTROSAT observation (left panel) and {\it GALEX} observation (right panel). While the lower panel visualize the NUV images of NGC~1316 using ASTROSAT observation (left panel) and {\it GALEX} observation (right panel)..}
\label{fig1}
\end{figure*}

\section{Observations and Analysis}

The first Indian multi-wavelength astronomical satellite {\it AstroSat} was launched by Indian Space Research Organization (ISRO) on 28 September 2015 (Agrawal 2006, Singh et al. 2014). The observatory is sensitive to photons from visible, UV and X-ray bands simultaneously through five sets of instruments such as Ultraviolet Imaging Telescope (UVIT; Tandon et al. 2017), Soft X-ray Telescope (SXT; Singh et al. 2017), Large Area X-ray Proportional Counters (LAXPCs; Agrawal et al. 2017), Cadmium Zinc Telluride Imager (CZTI; Rao et al.  2017), and a Scanning Sky Monitor (SSM; Ramadevi et al. 2018). The UVIT consists of two co-aligned telescopes with diameters of 38 cm, in Ritchey-Chretien configuration. Out of the two, one telescope is fully dedicated for FUV observations, whereas the other one observes in NUV and visible channels. The UV imaging telescope covers $\sim$28$^{\prime}$ circular field of view with an angular resolution of $\approx 1.2$\arcs\, for the NUV and $< 1.4$\arcs\, for FUV channels. The details of UV imaging telescope and instrumentation can be found in \citet{2017JApA...38...28T,2017AJ....154..128T}. 

The UVIT on-board {\it AstroSat} was used to observe our target galaxy, NGC~1316 at a redshift located at a luminosity  distance\footnote{\url{http://www.astro.ucla.edu/\%7Ewright/CosmoCalc.html}} \s 25 Mpc \citep{2006PASP..118.1711W}. The angular scale obtained by angular size distance of 1\arcs corresponds to 0.122 kpc. The observations of the galaxy NGC~1316 were carried out in four UVIT filters. The characteristics of the filters, their mean wavelength and band width, exposure time, zero point and unit conversion factors are tabulated in Table~\ref{tab2}. Photometric calibrations for FUV and NUV filters were carried out by following the procedure described in \citet{2017JApA...38...28T,2017AJ....154..128T}. 
 
\begin{table*}[htb]
\centering
\tabularfont
\caption{Details of the observation of NGC~1316}\label{tab2} 
\begin{tabular}{lllllcc}
\topline 
Filter(slot)	 &Name	        & mean($\lambda$)	& $\Delta \lambda $	&Zero-Point   &Unit conversion (UC)  &Exp. Time\\
		 &		        & (in \A)                    & (in \A)        &		 & (\tim 10$\, ^{-15}$ $erg\, cm^{-2} s^{-1}$\A)   &(in second)\\
		 \midline  
\underline {FUV channel} & & & & & & \\
F148W (F1)   & CaF2-1     & 1481 			& 500 			& 18.016\pms0.01   & 3.09\pms0.029     & 830  \\
F154W (F2)   & BaF2	      & 1541 			& 380	    	& 17.778\pms0.01   & 3.55\pms0.04      & 300  \\
F169M (F3)   & Sapphire   & 1608      		& 290       	& 17.455\pms0.01   & 4.392\pms0.037     & 1280 \\
F172M (F5)   & Silica	  & 1717     		& 125       	& 16.342\pms0.02   & 10.74\pms0.16      & 874  \\
\\
\underline {NUV channel} &  & & & & & \\
N219M (F2)  & NUVB15      & 2196			&  270       	& 16.50\pms0.01   & 3.50\pms0.035      & 3747 \\
\hline
\end{tabular}
\end{table*}

\section{Imaging Analysis}

The high resolution UVIT data on NGC~1316 acquired using {\it AstroSat} was reduced following standard routine described in \citet{2017AJ....154..128T}. The level~2 image files were obtained from level~1 data by applying the {\it UVIT Level-2 Pipeline} (UL2P) task. This consists of the preliminary reduction, masking of bad pixels and flagging of multi-photon event blocks, detection of cosmic-ray events that affect the science data and finally applying the drift correction to the event centroids. Then, the multiple orbital image files were combined after applying astrometric corrections delivered by the \texttt{UVIT\_DriverModule} task. The final images thus produced were the intensity maps in units of counts per second. This paper uses data sets acquired using two UV bands, namely NUV (N219M) and FUV (F148W) while data in other bands were rejected due to the poor signal.

The flux and AB magnitude measurements of NGC~1316 were obtained by multiplying the count rates in a particular filter by its corresponding conversion factor as given by \citet{2017JApA...38...28T}. The source and background counts were extracted using {\it imcnts} task within {\it xspatial} package of Image Reduction and Analysis Facility ({\tt IRAF}). The background counts were extracted by selecting multiple source free regions in image files. Then the counts in the science image were corrected for background emission and normalized considering the integration time. 

\begin{figure}
\centering\includegraphics[width=0.45\textwidth]{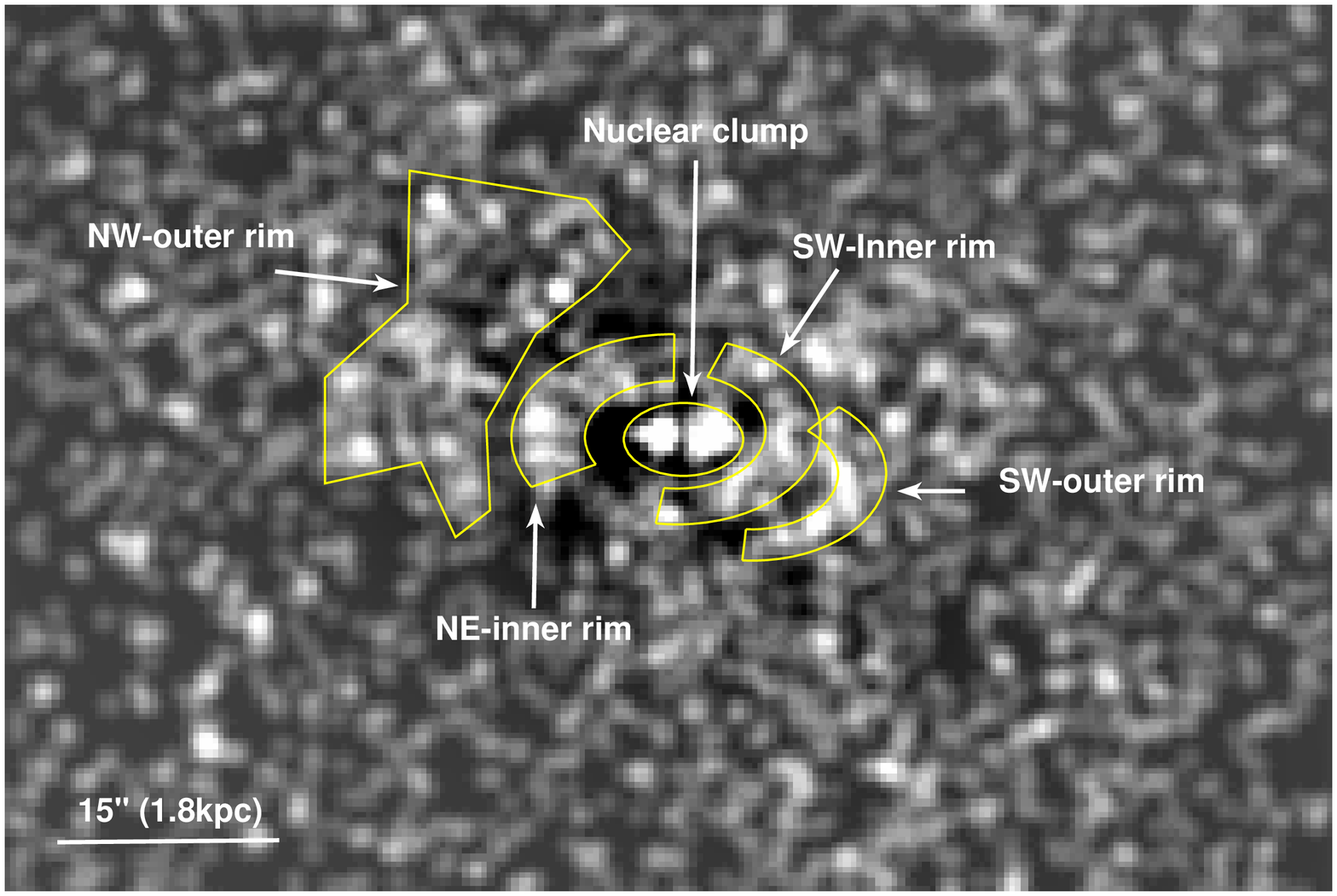}
\centering\includegraphics[width=0.45\textwidth]{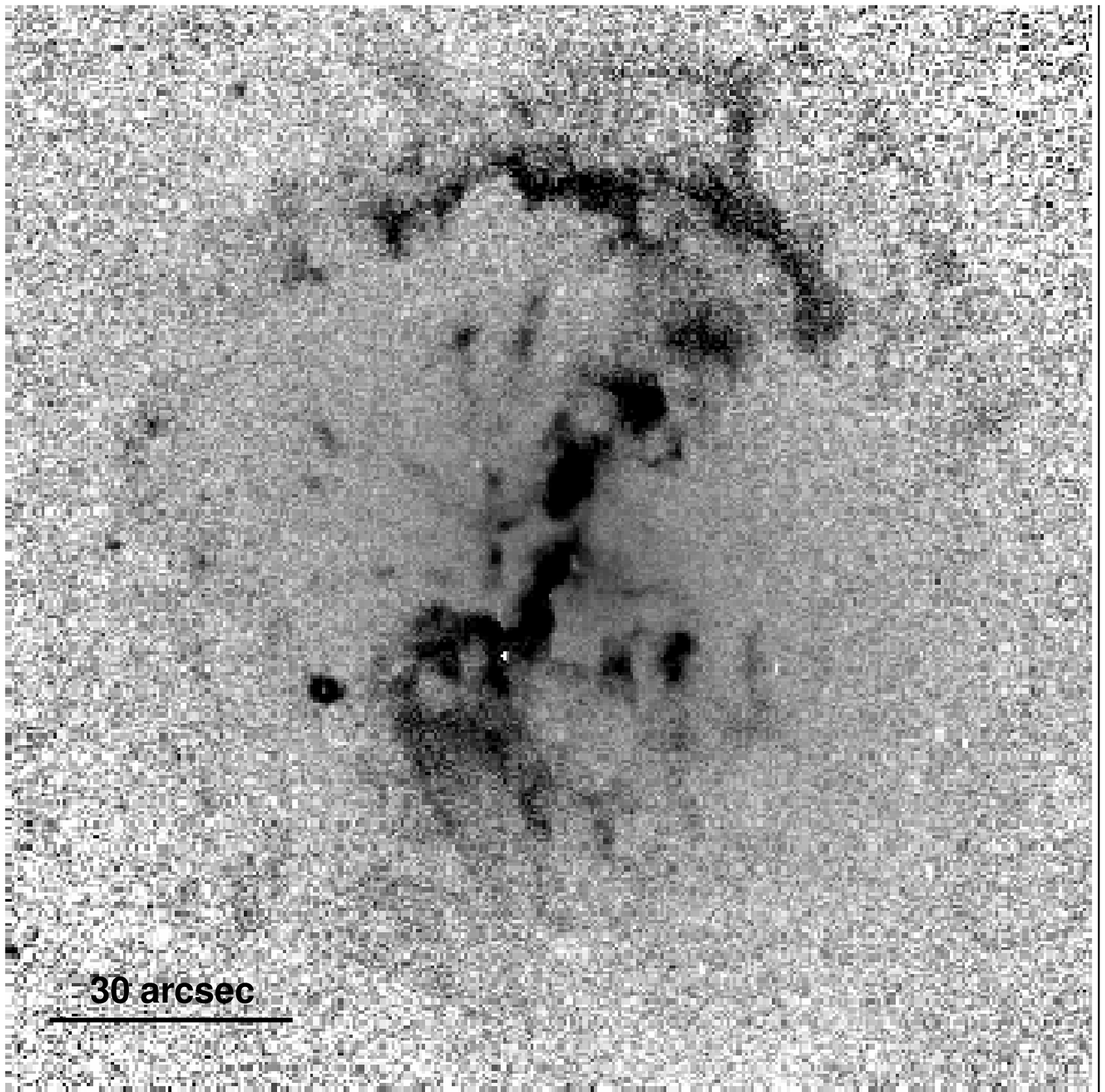}
\centering\includegraphics[width=0.45\textwidth]{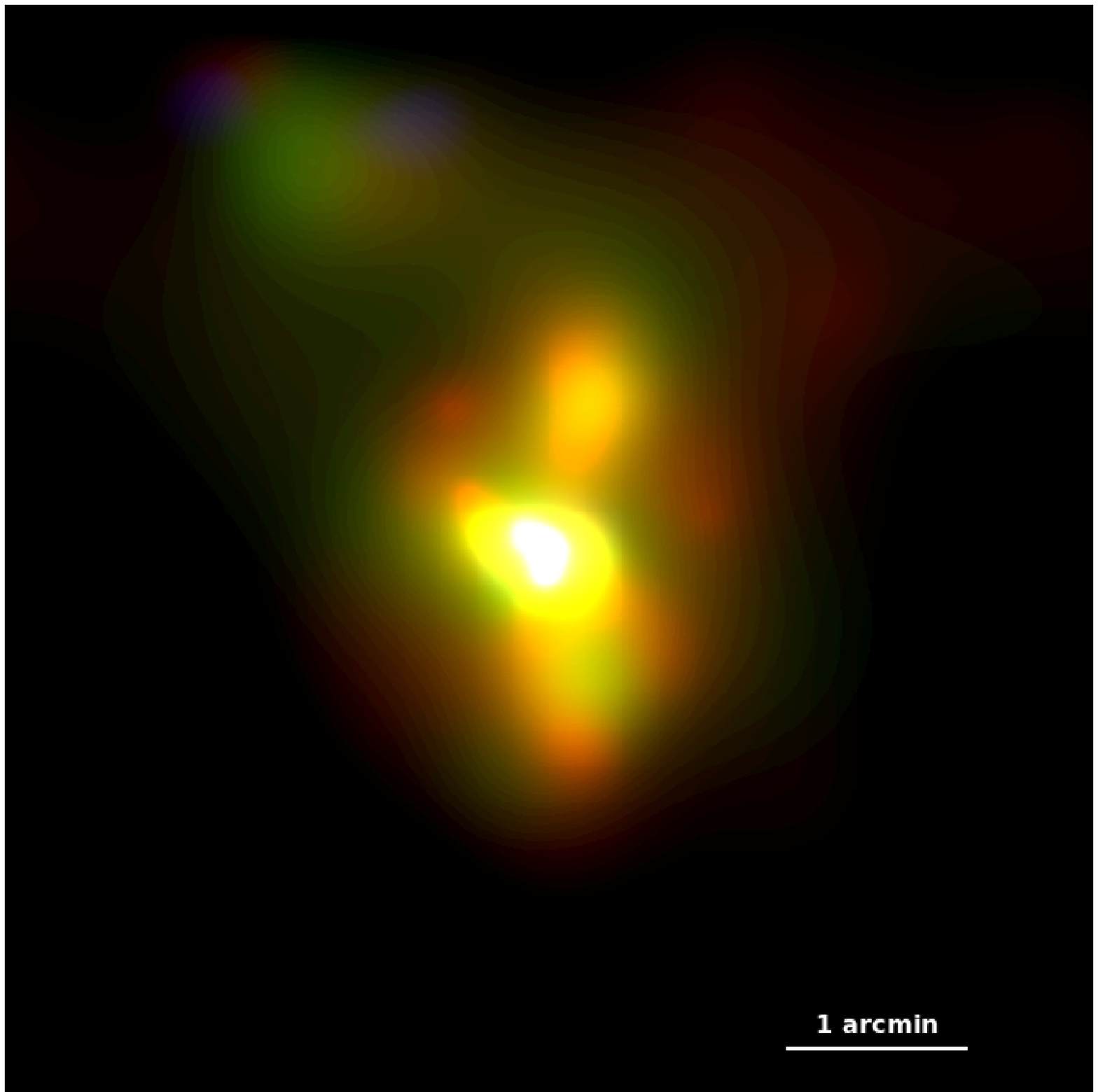}
\caption{{\it Top panel shows} the {\it AstroSat}/UVIT FUV(F148W) residual image of NGC~1316, created by subtracting the model image from the original background subtracted raw image. Arrow marks in the image point to the excess emission from the central region. The FUV emission directly represents the location of young massive stars in the nebula. {\it middle panel} represents the \bv colour map and {\it bottom panel} shows the distribution of hot gas.}
\label{fig2}
\end{figure}

The astrometry corrected and 1.5$\sigma$ Gaussian smoothed FUV and NUV emission maps, derived from the analysis of the {\it AstroSat} data are shown in the left panels of Figure~\ref{fig1}, while those in the right panels compare them with that of the data from {\it GALEX}. The small scale features apparent within few arcsec ($\approx$5\arcs) in {\it AstroSat} FUV data are not detected by {\it GALEX}. Similar small scale structures are also evident in the {\it AstroSat} NUV images. The imaging comparison of the data from {\it AstroSat} and {\it GALEX} observatories find the extent of FUV emission up to \s 20\arcs\, against \s 70\arcs\ in {\it GALEX}, while that in NUV bands is found to be \s 30\arcs\, against \s 2\arcm\, in {\it GALEX}. To enhance the visibility of the UV emission features in NGC~1316, the residual maps presented in Figure~\ref{fig2} are derived after subtracting the smooth models in FUV and NUV bands of {\it AstroSat} data. Here, smooth models of the FUV and NUV emission were obtained by fitting ellipses using task {\it ellipse} within {\tt IRAF~2.16} to the surface brightness distribution in the respective bands. This was achieved by obtaining its smooth models after fitting ellipses to the isophotes of the galaxy image using \texttt{ellipse} task within {\tt IRAF} (for details see \citet{2007A&A...461..103P}; \citet{2012NewA...17..524V}). Smooth 2D models thus derived which were then subtracted from the cleaned science frames of NGC~1316. One of the residual maps of NGC~1316 in FUV is shown in Figure~\ref{fig2}, which confirms the presence of the hidden features like knots, multiple rim-like structures along the NE and SW directions and a depression in the surface brightness near its nucleus (within central 1-2 kpc) and collectively point towards the complex morphological structures of NGC~1316. With an objective to investigate the association of the apparent UV extinction, we have made use of the optical B and V band data sets acquired using CTIO 1.5 m telescope and have generated the (B-V) colour index map of NGC~1316. The (B-V) colour index map reveals very complex dust extinction regions (darker shades) oriented along the optical minor axis of NGC~1316 in the form of a lane, which then appears in the form of a complex arc shape on either side of the lane. Dust lane, apparent in the central region of (B-V) colour index map, is orthogonal to its UV emission. However, the UV emission appears to be absorbed at the locations of dust in the central region of NGC~1316. We also generated a map of hot X-ray emitting gas distribution within NGC~1316 from the analysis of high resolution 0.3 to 6 keV {\tt Chandra} data and is shown in the lower panel of Figure~\ref{fig2}. X-ray emission from this galaxy also exhibit a complex structure, however, by and large the diffuse X-ray emission appears to be oriented along dust lane extended to even larger extent ($\sim$2 arcmin).

\begin{figure*}
\centering\includegraphics[width=0.85\textwidth]{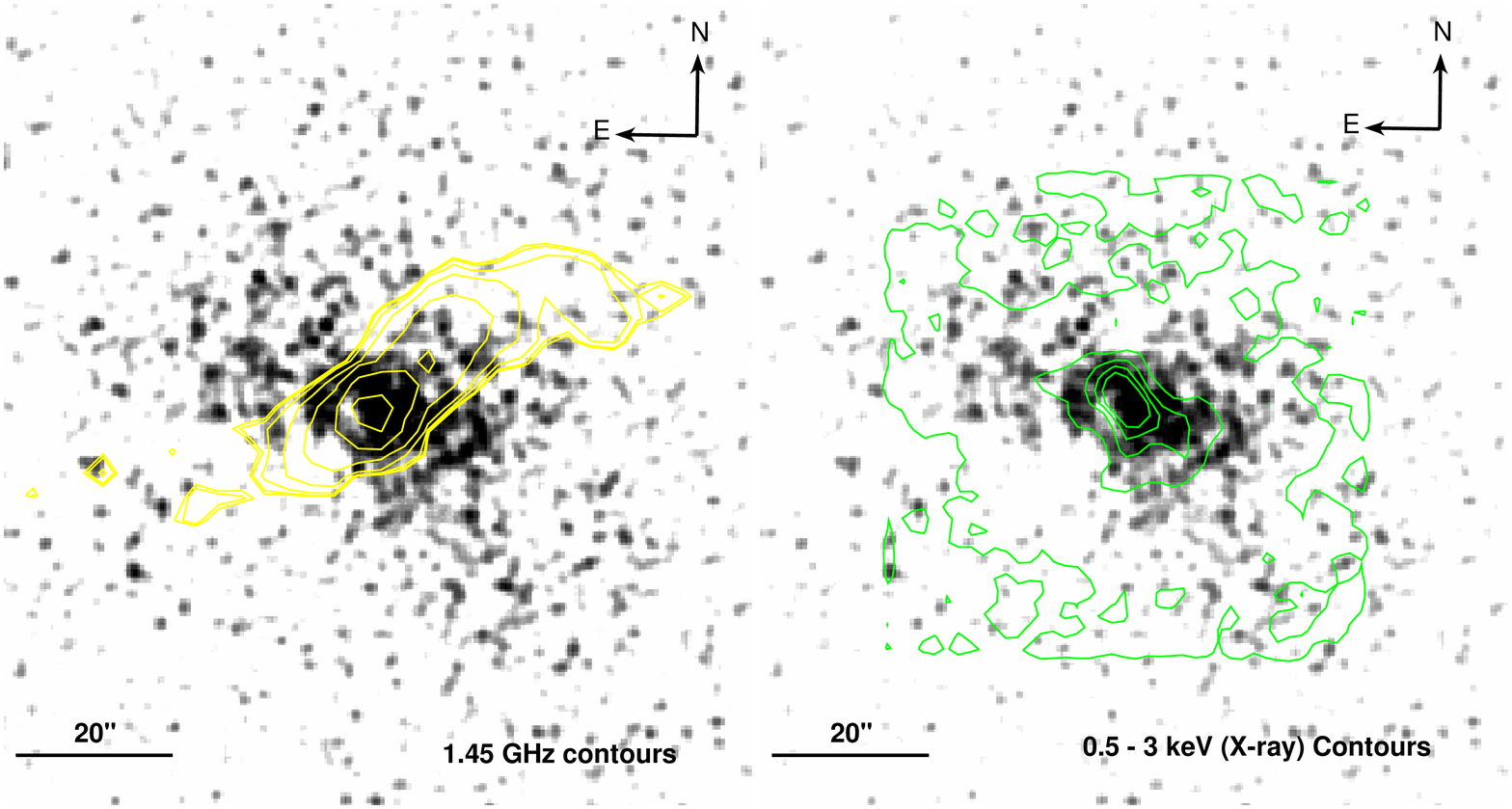}
\caption{The FUV (F148W) images are Gaussian smoothed with 1.5\arcs. The left panel shows the 1.45 GHz radio contours overlaid on the FUV image, whereas the right panel shows the soft X-ray contours overlaid on the FUV image. }
\label{fig3}
\end{figure*}

\section{Discussion}

The rate at which the galaxies form stars is only one of the several fundamental astrophysical processes in galaxy evolution. Massive young O, B or A-type stars are hot and emit a quantitative amount of ultraviolet radiation and thus exhibit an ideal tracer to identify and measure the star formation activity of the galaxy. The space-based telescopes such as {\it AstroSat} and {\it GALEX} provide direct estimate of the current star-formation rate (SFR) by measuring the UV radiation. The star formation rate of NGC~1316 was calculated by measuring the luminosity (or flux) in NUV and FUV bands. The total measured flux within 20\arcs\, in FUV and NUV are (24.0\pms8.0)\,\tim 10$^{-12}$~\flux and (38.8\pms8.5)\,\tim 10$^{-12}$~\flux , respectively. This corresponds to the luminosity of (1.85\pms0.62) \tim 10$^{42}$ \lum\, and (2.99\pms0.66) \tim 10$^{42}$ \lum\,, respectively. The SFR in the galaxy were estimated from its UV luminosity and employing the relation given by \citet{2006ApJS..164...38I} which can be expressed as,
\begin{align*}
    SFR\, (FUV)\,(M_{\odot} yr^{-1}) & = \frac{L_{(FUV)}\,(erg\, s^{-1})}{3.83 \times 10^{33}} \times 10^{-9.51} \\
     SFR\, (NUV)\,(M_{\odot} yr^{-1}) & = \frac{L_{(NUV)}\,(erg\, s^{-1})}{3.83 \times 10^{33}} \times 10^{-9.33} 
\end{align*}

Using the estimated FUV and NUV luminosities from the UVIT observation, the star formation rates of NGC~1316 are, respectively, found to be 0.15\pms0.05 \Msun \pyr and 0.36\pms0.07  \Msun \pyr and provide lower limit due to the foreground screening effect. But the apparent significant and complex dust extinction region imply that the UV emission from this galaxy is extinguished and therefore the SFR estimated using the UV fluxes alone represent a substantial underestimate. As a result, other possible tracers need to be explored for better representation. The UV photons absorbed by the dust grain produces re-emission at longer wavelengths like IR or FIR and hence IR emission analysis is called for this system. \citet{2016A&A...592A..20D} using 22 $\mu m$ luminosity derived the star-formation rate of  \s 0.7  \Msun\pyr, whereas the IRAS FIR luminosities yielded the SFR as 0.82  \Msun \pyr \citep{2017ApJ...844..170T}.

NGC~1316 is a well-known merger type galaxy and merger induces star formation at the centre (100-1000 pc) of the galaxy \citep{2011EAS....51..107B}. The nuclear activity in merger-induced starburst galaxies have long been given more importance in the study of merger and star formation connection \citep{1988ApJ...325...74S,1997A&AS..124..533D,1989MNRAS.240..329L}. Though, NGC~1316 satisfies characteristics of merger remnant, gas-rich system, the SFRs in this galaxy are significantly less. Understanding the cause of quenching of star formation in NGC~1316 is, therefore, a challenge for studies of evolution of the galaxy. A different theoretical explanation has been proposed for this challenge though a general agreement is yet to reach. The most acceptable explanation is the Active Galactic Nucleus (AGN) outflows can remove the substantial amount of gas from the host galaxy, thereby shutting off the star formation. On the other hand, these outflows compress the surrounding gas and trigger star formation in compressed outer region. To investigate the outflow activity, we used radio image of NGC~1316, observed on 31 May 2002 (ObsID: AH787) available in VLA archive. The {\it left panel} of Figure~\ref{fig3} shows the FUV image of NGC~1316 along with the 1.45 GHz VLA radio contours. The direction of the radio jet can be seen along the FUV surface brightness depression region implying that the radio jet pushes (remove) the gas in outward direction from the central region of NGC~1316. From this map it is also apparent that some amount of gas gets compressed along the NW and SE direction and emits radiation in the ultraviolet region. Thus, the AGN driven outflows remove gas from the nuclear region and suppresses star formation in the central region. 

The NUV and FUV emission maps derived from the analysis of ASTROSAT data exhibit multiple rim-like structures along the NE and SW directions and a significant depression in the central region of surface brightness from NGC~1316. The presence of such peculiar structures in NGC~1316 is likely to be due to the tidal forces that pulled gas and stars form core after the merger episode or due to the AGN outburst. The radio jets originating from central engine pushes away the gas in outward direction during outbursts \citep{2011ApJ...735...11O,2009ApJ...705..624D,2000ApJ...534L.135M,2007MNRAS.381.1381S,2019MNRAS.485.1981V}. However, the recent high resolution radio observations performed using MeerKAT suggests its origin to be due to a $\sim$ 10:1 merger between a dominant early-type galaxy and a smaller, gas-rich progenitor \citep{2019PASJ...71...85M}.

Spatial correspondence between the FUV, 0.5--3 keV X-ray and 1.45 GHz radio emission from NGC~1316 are shown in Figure~\ref{fig3}. X-ray emission map was obtained using high resolution {\it Chandra} 30 ks archival data (Obs. ID 2022) with the ACIS-S3 as the target. The left panel of Figure~\ref{fig3} shows 1.4 GHz radio contours overlaid on the FUV image of NGC~1316, while in the right panel, the 0.5-3 keV X-ray contours are overlaid on the FUV image. The figure clearly shows the morphological similarity in the FUV and X-ray emission. Both these indicate deficient regions along the NE and SW directions, whereas excess emission is apparent in both the bands along the NW and SE directions pointing towards their common origin. It is likely that the observed deficiencies are carved by the central AGN, radio jets of which pushes the UV and X-ray emitting gas from the central region of the galaxy.

\section{Conclusion} 

We have observed the nearby merger remnant galaxy NGC~1316 in search of the connection between nuclear activity and star formation in its central region using high resolution NUV and FUV imaging telescopes on board the first Indian dedicated astronomical satellite {\it AstroSat}. The unsharp masked image as well as the surface brightness profiles from the analysis of these data sets confirm that the UV emission from this galaxy is not smooth but exhibits perturbations and are in agreement with the results presented by several other researchers. Some of the important results from this study are:

\begin{itemize}
    \item The residual image confirms the presence of peculiar features in the nuclear (1-2 kpc) region of NGC~1316.
     
    \item The hidden structures like rim, clumps and their strong spatial correspondence with imagery at other wavelengths confirm that the origin of gas and dust in this system is due to the merger like episodes.
    
    \item The estimates of star formation rates in NGC~1316 based on FUV and NUV luminosities are 0.15 \Msun\pyr and 0.36 \Msun\pyr, respectively.
    
    \item A comparison of 1.4 GHz emission contours and FUV and emission map revealed that the UV emitting sources are displaced away by the radio jets emanating from the central engine, thereby confirming that the AGN driven outflows are responsible for the quenching of the star formation in the gas-rich merger remnant galaxy NGC~1316.
\end{itemize}

\section*{Acknowledgements}
We thank the reviewer for encouraging comments and suggestions on the paper. This work uses the data from the AstroSat mission of the Indian Space Research Organisation (ISRO), archived at the Indian Space Science Data Centre (ISSDC). The UVIT project is a result of the collaboration between IIA, Bengaluru, IUCAA, Pune, TIFR, Mumbai, several centers of ISRO, and CSA. Data from VLA and Chandra archive, Extragalactic Database (NED) and NASA's Astrophysics Data System (ADS) are also used in this paper. NDV thanks Science and Engineering Research Board (SERB), India for providing a research fund (Ref. No.: YSS/2015/001413). NDV also thanks IUCAA, Pune for the use of library facility.
\vspace{-1em}

\def\aj{AJ}%
\def\actaa{Acta Astron.}%
\def\araa{ARA\&A}%
\def\apj{ApJ}%
\def\apjl{ApJ}%
\def\apjs{ApJS}%
\def\ao{Appl.~Opt.}%
\def\apss{Ap\&SS}%
\def\aap{A\&A}%
\def\aapr{A\&A~Rev.}%
\def\aaps{A\&AS}%
\def\azh{AZh}%
\def\baas{BAAS}%
\def\bac{Bull. astr. Inst. Czechosl.}%
\def\caa{Chinese Astron. Astrophys.}%
\def\cjaa{Chinese J. Astron. Astrophys.}%
\def\icarus{Icarus}%
\def\jcap{J. Cosmology Astropart. Phys.}%
\def\jrasc{JRASC}%
\def\mnras{MNRAS}%
\def\memras{MmRAS}%
\def\na{New A}%
\def\nar{New A Rev.}%
\def\pasa{PASA}%
\def\pra{Phys.~Rev.~A}%
\def\prb{Phys.~Rev.~B}%
\def\prc{Phys.~Rev.~C}%
\def\prd{Phys.~Rev.~D}%
\def\pre{Phys.~Rev.~E}%
\def\prl{Phys.~Rev.~Lett.}%
\def\pasp{PASP}%
\def\pasj{PASJ}%
\def\qjras{QJRAS}%
\def\rmxaa{Rev. Mexicana Astron. Astrofis.}%
\def\skytel{S\&T}%
\def\solphys{Sol.~Phys.}%
\def\sovast{Soviet~Ast.}%
\def\ssr{Space~Sci.~Rev.}%
\def\zap{ZAp}%
\def\nat{Nature}%
\def\iaucirc{IAU~Circ.}%
\def\aplett{Astrophys.~Lett.}%
\def\apspr{Astrophys.~Space~Phys.~Res.}%
\def\bain{Bull.~Astron.~Inst.~Netherlands}%
\def\fcp{Fund.~Cosmic~Phys.}%
\def\gca{Geochim.~Cosmochim.~Acta}%
\def\grl{Geophys.~Res.~Lett.}%
\def\jcp{J.~Chem.~Phys.}%
\def\jgr{J.~Geophys.~Res.}%
\def\jqsrt{J.~Quant.~Spec.~Radiat.~Transf.}%
\def\memsai{Mem.~Soc.~Astron.~Italiana}%
\def\nphysa{Nucl.~Phys.~A}%
\def\physrep{Phys.~Rep.}%
\def\physscr{Phys.~Scr}%
\def\planss{Planet.~Space~Sci.}%
\def\procspie{Proc.~SPIE}%
\let\astap=\aap
\let\apjlett=\apjl
\let\apjsupp=\apjs
\let\applopt=\ao
\bibliography{mybib}

\begin{thebibliography}{}
\expandafter\ifx\csname natexlab\endcsname\relax\def\natexlab#1{#1}\fi

\bibitem[{{Bournaud}(2011)}]{2011EAS....51..107B}
{Bournaud}, F. 2011, in EAS Publications Series, Vol.~51, EAS Publications
  Series, ed. C.~{Charbonnel} \& T.~{Montmerle}, 107--131

\bibitem[{{Buat} {$et~al$.}(2010){Buat}, {Giovannoli}, {Burgarella}, {Altieri},
  {Amblard}, {Arumugam}, {Aussel}, {Babbedge}, {Blain}, {Bock}, {Boselli},
  {Castro-Rodr{\'\i}guez}, {Cava}, {Chanial}, {Clements}, {Conley}, {Conversi},
  {Cooray}, {Dowell}, {Dwek}, {Eales}, {Elbaz}, {Fox}, {Franceschini}, {Gear},
  {Glenn}, {Griffin}, {Halpern}, {Hatziminaoglou}, {Heinis}, {Ibar}, {Isaak},
  {Ivison}, {Lagache}, {Levenson}, {Lonsdale}, {Lu}, {Madden}, {Maffei},
  {Magdis}, {Mainetti}, {Marchetti}, {Morrison}, {Nguyen}, {O'Halloran},
  {Oliver}, {Omont}, {Owen}, {Page}, {Pannella}, {Panuzzo}, {Papageorgiou},
  {Pearson}, {P{\'e}rez-Fournon}, {Pohlen}, {Rigopoulou}, {Rizzo}, {Roseboom},
  {Rowan-Robinson}, {S{\'a}nchez Portal}, {Schulz}, {Seymour}, {Shupe},
  {Smith}, {Stevens}, {Strazzullo}, {Symeonidis}, {Trichas}, {Tugwell},
  {Vaccari}, {Valiante}, {Valtchanov}, {Vigroux}, {Wang}, {Ward}, {Wright},
  {Xu}, \& {Zemcov}}]{2010MNRAS.409L...1B}
{Buat}, V., {Giovannoli}, E., {Burgarella}, D., {$et~al$.} 2010, \mnras, 409,
  L1

\bibitem[{{Carlqvist}(2010)}]{2010Ap&SS.327..267C}
{Carlqvist}, P. 2010, \apss, 327, 267

\bibitem[{{Croton} {$et~al$.}(2006){Croton}, {Springel}, {White}, {De Lucia},
  {Frenk}, {Gao}, {Jenkins}, {Kauffmann}, {Navarro}, \&
  {Yoshida}}]{2006MNRAS.365...11C}
{Croton}, D.~J., {Springel}, V., {White}, S. D.~M., {$et~al$.} 2006, \mnras,
  365, 11

\bibitem[{{David} {$et~al$.}(2009){David}, {Jones}, {Forman}, {Nulsen},
  {Vrtilek}, {O'Sullivan}, {Giacintucci}, \&
  {Raychaudhury}}]{2009ApJ...705..624D}
{David}, L.~P., {Jones}, C., {Forman}, W., {$et~al$.} 2009, \apj, 705, 624

\bibitem[{{Deshmukh} {$et~al$.}(2013){Deshmukh}, {Tate}, {Vagshette}, {Pandey},
  \& {Patil}}]{2013RAA....13..885D}
{Deshmukh}, S.~P., {Tate}, B.~T., {Vagshette}, N.~D., {Pandey}, S.~K., \&
  {Patil}, M.~K. 2013, Research in Astronomy and Astrophysics, 13, 885

\bibitem[{{Draine} {$et~al$.}(2007){Draine}, {Dale}, {Bendo}, {Gordon},
  {Smith}, {Armus}, {Engelbracht}, {Helou}, {Kennicutt}, {Li}, {Roussel},
  {Walter}, {Calzetti}, {Moustakas}, {Murphy}, {Rieke}, {Bot}, {Hollenbach},
  {Sheth}, \& {Teplitz}}]{2007ApJ...663..866D}
{Draine}, B.~T., {Dale}, D.~A., {Bendo}, G., {$et~al$.} 2007, \apj, 663, 866

\bibitem[{{Drinkwater} {$et~al$.}(2001){Drinkwater}, {Gregg}, {Holman}, \&
  {Brown}}]{2001MNRAS.326.1076D}
{Drinkwater}, M.~J., {Gregg}, M.~D., {Holman}, B.~A., \& {Brown}, M.~J.~I.
  2001, \mnras, 326, 1076

\bibitem[{{Duah Asabere} {$et~al$.}(2016){Duah Asabere}, {Horellou}, {Jarrett},
  \& {Winkler}}]{2016A&A...592A..20D}
{Duah Asabere}, B., {Horellou}, C., {Jarrett}, T.~H., \& {Winkler}, H. 2016,
  \aap, 592, A20

\bibitem[{{Duc} {$et~al$.}(1997){Duc}, {Mirabel}, \&
  {Maza}}]{1997A&AS..124..533D}
{Duc}, P.~A., {Mirabel}, I.~F., \& {Maza}, J. 1997, \aaps, 124, 533

\bibitem[{{Ekers} {$et~al$.}(1983){Ekers}, {Goss}, {Wellington}, {Bosma},
  {Smith}, \& {Schweizer}}]{1983A&A...127..361E}
{Ekers}, R.~D., {Goss}, W.~M., {Wellington}, K.~J., {$et~al$.} 1983, \aap, 127,
  361

\bibitem[{{Fabian}(2012)}]{2012ARA&A..50..455F}
{Fabian}, A.~C. 2012, \araa, 50, 455

\bibitem[{{Goudfrooij} {$et~al$.}(2001){Goudfrooij}, {Alonso}, {Maraston}, \&
  {Minniti}}]{2001MNRAS.328..237G}
{Goudfrooij}, P., {Alonso}, M.~V., {Maraston}, C., \& {Minniti}, D. 2001,
  \mnras, 328, 237

\bibitem[{{Hopkins} {$et~al$.}(2008){Hopkins}, {McClure-Griffiths}, \&
  {Gaensler}}]{2008ApJ...682L..13H}
{Hopkins}, A.~M., {McClure-Griffiths}, N.~M., \& {Gaensler}, B.~M. 2008, \apjl,
  682, L13

\bibitem[{{Horellou} {$et~al$.}(2001){Horellou}, {Black}, {van Gorkom},
  {Combes}, {van der Hulst}, \& {Charmandaris}}]{2001A&A...376..837H}
{Horellou}, C., {Black}, J.~H., {van Gorkom}, J.~H., {$et~al$.} 2001, \aap,
  376, 837

\bibitem[{{Iglesias-P{\'a}ramo} {$et~al$.}(2006){Iglesias-P{\'a}ramo}, {Buat},
  {Takeuchi}, {Xu}, {Boissier}, {Boselli}, {Burgarella}, {Madore}, {Gil de
  Paz}, {Bianchi}, {Barlow}, {Byun}, {Donas}, {Forster}, {Friedman}, {Heckman},
  {Jelinski}, {Lee}, {Malina}, {Martin}, {Milliard}, {Morrissey}, {Neff},
  {Rich}, {Schiminovich}, {Seibert}, {Siegmund}, {Small}, {Szalay}, {Welsh}, \&
  {Wyder}}]{2006ApJS..164...38I}
{Iglesias-P{\'a}ramo}, J., {Buat}, V., {Takeuchi}, T.~T., {$et~al$.} 2006,
  \apjs, 164, 38

\bibitem[{Iodice {$et~al$.}(2017)Iodice, Spavone, Capaccioli, Peletier,
  Richtler, Hilker, Mieske, Limatola, Grado, Napolitano, Cantiello, D'Abrusco,
  Paolillo, Venhola, Lisker, de~Ven, Falcon-Barroso, \&
  Schipani}]{2017ApJ...839...21I}
Iodice, E., Spavone, M., Capaccioli, M., {$et~al$.} 2017, \apj, 839, 21

\bibitem[{{Kennicutt} \& {Evans}(2012)}]{2012ARA&A..50..531K}
{Kennicutt}, R.~C., \& {Evans}, N.~J. 2012, \araa, 50, 531

\bibitem[{{Lanz} {$et~al$.}(2010){Lanz}, {Jones}, {Forman}, {Ashby}, {Kraft},
  \& {Hickox}}]{2010ApJ...721.1702L}
{Lanz}, L., {Jones}, C., {Forman}, W.~R., {$et~al$.} 2010, \apj, 721, 1702

\bibitem[{{Lawrence} {$et~al$.}(1989){Lawrence}, {Rowan-Robinson}, {Leech},
  {Jones}, \& {Wall}}]{1989MNRAS.240..329L}
{Lawrence}, A., {Rowan-Robinson}, M., {Leech}, K., {Jones}, D.~H.~P., \&
  {Wall}, J.~V. 1989, \mnras, 240, 329

\bibitem[{{Malin} \& {Carter}(1983)}]{1983ApJ...274..534M}
{Malin}, D.~F., \& {Carter}, D. 1983, \apj, 274, 534

\bibitem[{{Mancuso} {$et~al$.}(2017){Mancuso}, {Lapi}, {Prandoni}, {Obi},
  {Gonzalez-Nuevo}, {Perrotta}, {Bressan}, {Celotti}, \&
  {Danese}}]{2017ApJ...842...95M}
{Mancuso}, C., {Lapi}, A., {Prandoni}, I., {$et~al$.} 2017, \apj, 842, 95

\bibitem[{{McKinley} {$et~al$.}(2015){McKinley}, {Yang}, {L{\'o}pez-Caniego},
  {Briggs}, {Hurley-Walker}, {Wayth}, {Offringa}, {Crocker}, {Bernardi},
  {Procopio}, {Gaensler}, {Tingay}, {Johnston-Hollitt}, {McDonald}, {Bell},
  {Bhat}, {Bowman}, {Cappallo}, {Corey}, {Deshpande}, {Emrich}, {Ewall-Wice},
  {Feng}, {Goeke}, {Greenhill}, {Hazelton}, {Hewitt}, {Hindson}, {Jacobs},
  {Kaplan}, {Kasper}, {Kratzenberg}, {Kudryavtseva}, {Lenc}, {Lonsdale},
  {Lynch}, {McWhirter}, {Mitchell}, {Morales}, {Morgan}, {Oberoi}, {Ord},
  {Pindor}, {Prabu}, {Riding}, {Rogers}, {Roshi}, {Udaya Shankar}, {Srivani},
  {Subrahmanyan}, {Waterson}, {Webster}, {Whitney}, {Williams}, \&
  {Williams}}]{2015MNRAS.446.3478M}
{McKinley}, B., {Yang}, R., {L{\'o}pez-Caniego}, M., {$et~al$.} 2015, \mnras,
  446, 3478

\bibitem[{{McNamara} {$et~al$.}(2000){McNamara}, {Wise}, {Nulsen}, {David},
  {Sarazin}, {Bautz}, {Markevitch}, {Vikhlinin}, {Forman}, {Jones}, \&
  {Harris}}]{2000ApJ...534L.135M}
{McNamara}, B.~R., {Wise}, M., {Nulsen}, P.~E.~J., {$et~al$.} 2000, \apjl, 534,
  L135

\bibitem[{{Morokuma-Matsui} {$et~al$.}(2019){Morokuma-Matsui}, {Serra},
  {Maccagni}, {For}, {Wang}, {Bekki}, {Morokuma}, {Egusa}, {Espada}, {Miura},
  {Nakanishi}, {Koribalski}, \& {Takeuchi}}]{2019PASJ...71...85M}
{Morokuma-Matsui}, K., {Serra}, P., {Maccagni}, F.~M., {$et~al$.} 2019, \pasj,
  71, 85

\bibitem[{{Oogi} \& {Habe}(2013)}]{2013MNRAS.428..641O}
{Oogi}, T., \& {Habe}, A. 2013, \mnras, 428, 641

\bibitem[{{Oogi} {$et~al$.}(2016){Oogi}, {Habe}, \&
  {Ishiyama}}]{2016MNRAS.456..300O}
{Oogi}, T., {Habe}, A., \& {Ishiyama}, T. 2016, \mnras, 456, 300

\bibitem[{{O'Sullivan} {$et~al$.}(2011){O'Sullivan}, {Giacintucci}, {David},
  {Gitti}, {Vrtilek}, {Raychaudhury}, \& {Ponman}}]{2011ApJ...735...11O}
{O'Sullivan}, E., {Giacintucci}, S., {David}, L.~P., {$et~al$.} 2011, \apj,
  735, 11

\bibitem[{{Pandge} {$et~al$.}(2012){Pandge}, {Vagshette}, {David}, \&
  {Patil}}]{2012MNRAS.421..808P}
{Pandge}, M.~B., {Vagshette}, N.~D., {David}, L.~P., \& {Patil}, M.~K. 2012,
  \mnras, 421, 808

\bibitem[{{Patil} {$et~al$.}(2007){Patil}, {Pandey}, {Sahu}, \&
  {Kembhavi}}]{2007A&A...461..103P}
{Patil}, M.~K., {Pandey}, S.~K., {Sahu}, D.~K., \& {Kembhavi}, A. 2007, \aap,
  461, 103

\bibitem[{{Rodriguez-Gomez} {$et~al$.}(2016){Rodriguez-Gomez}, {Pillepich},
  {Sales}, {Genel}, {Vogelsberger}, {Zhu}, {Wellons}, {Nelson}, {Torrey},
  {Springel}, {Ma}, \& {Hernquist}}]{2016MNRAS.458.2371R}
{Rodriguez-Gomez}, V., {Pillepich}, A., {Sales}, L.~V., {$et~al$.} 2016,
  \mnras, 458, 2371

\bibitem[{{Salim} {$et~al$.}(2007){Salim}, {Rich}, {Charlot}, {Brinchmann},
  {Johnson}, {Schiminovich}, {Seibert}, {Mallery}, {Heckman}, {Forster},
  {Friedman}, {Martin}, {Morrissey}, {Neff}, {Small}, {Wyder}, {Bianchi},
  {Donas}, {Lee}, {Madore}, {Milliard}, {Szalay}, {Welsh}, \&
  {Yi}}]{2007ApJS..173..267S}
{Salim}, S., {Rich}, R.~M., {Charlot}, S., {$et~al$.} 2007, \apjs, 173, 267

\bibitem[{{Sanders} {$et~al$.}(1988){Sanders}, {Soifer}, {Elias}, {Madore},
  {Matthews}, {Neugebauer}, \& {Scoville}}]{1988ApJ...325...74S}
{Sanders}, D.~B., {Soifer}, B.~T., {Elias}, J.~H., {$et~al$.} 1988, \apj, 325,
  74

\bibitem[{{Sanders} \& {Fabian}(2007)}]{2007MNRAS.381.1381S}
{Sanders}, J.~S., \& {Fabian}, A.~C. 2007, \mnras, 381, 1381

\bibitem[{{Schweizer}(1980)}]{1980ApJ...237..303S}
{Schweizer}, F. 1980, \apj, 237, 303

\bibitem[{{Schweizer} \& {Seitzer}(1988)}]{1988ApJ...328...88S}
{Schweizer}, F., \& {Seitzer}, P. 1988, \apj, 328, 88

\bibitem[{{Serra} {$et~al$.}(2019){Serra}, {Maccagni}, {Kleiner}, {de Blok},
  {van Gorkom}, {Hugo}, {Iodice}, {J{\'o}zsa}, {Kamphuis}, {Kraan-Korteweg},
  {Loni}, {Makhathini}, {Moln{\'a}r}, {Oosterloo}, {Peletier}, {Ramaila},
  {Ramatsoku}, {Smirnov}, {Smith}, {Spavone}, {Thorat}, {Trager}, \&
  {Venhola}}]{2019A&A...628A.122S}
{Serra}, P., {Maccagni}, F.~M., {Kleiner}, D., {$et~al$.} 2019, \aap, 628, A122

\bibitem[{{Sonkamble} {$et~al$.}(2015){Sonkamble}, {Vagshette}, {Pawar}, \&
  {Patil}}]{2015Ap&SS.359...61S}
{Sonkamble}, S.~S., {Vagshette}, N.~D., {Pawar}, P.~K., \& {Patil}, M.~K. 2015,
  \apss, 359, 21

\bibitem[{{Tandon} {$et~al$.}(2017{\natexlab{a}}){Tandon}, {Subramaniam},
  {Girish}, {Postma}, {Sankarasubramanian}, {Sriram}, {Stalin}, {Mondal},
  {Sahu}, {Joseph}, {Hutchings}, {Ghosh}, {Barve}, {George}, {Kamath},
  {Kathiravan}, {Kumar}, {Lancelot}, {Leahy}, {Mahesh}, {Mohan},
  {Nagabhushana}, {Pati}, {Kameswara Rao}, {Sreedhar}, \&
  {Sreekumar}}]{2017AJ....154..128T}
{Tandon}, S.~N., {Subramaniam}, A., {Girish}, V., {$et~al$.}
  2017{\natexlab{a}}, \aj, 154, 128

\bibitem[{{Tandon} {$et~al$.}(2017{\natexlab{b}}){Tandon}, {Hutchings},
  {Ghosh}, {Subramaniam}, {Koshy}, {Girish}, {Kamath}, {Kathiravan}, {Kumar},
  {Lancelot}, {Mahesh}, {Mohan}, {Murthy}, {Nagabhushana}, {Pati}, {Postma},
  {Rao}, {Sankarasubramanian}, {Sreekumar}, {Sriram}, {Stalin}, {Sutaria},
  {Sreedhar}, {Barve}, {Mondal}, \& {Sahu}}]{2017JApA...38...28T}
{Tandon}, S.~N., {Hutchings}, J.~B., {Ghosh}, S.~K., {$et~al$.}
  2017{\natexlab{b}}, Journal of Astrophysics and Astronomy, 38, 28

\bibitem[{{Terlevich} \& {Forbes}(2002)}]{2002MNRAS.330..547T}
{Terlevich}, A.~I., \& {Forbes}, D.~A. 2002, \mnras, 330, 547

\bibitem[{{Terrazas} {$et~al$.}(2017){Terrazas}, {Bell}, {Woo}, \&
  {Henriques}}]{2017ApJ...844..170T}
{Terrazas}, B.~A., {Bell}, E.~F., {Woo}, J., \& {Henriques}, B. M.~B. 2017,
  \apj, 844, 170

\bibitem[{{Vagshette} {$et~al$.}(2019){Vagshette}, {Naik}, \&
  {Patil}}]{2019MNRAS.485.1981V}
{Vagshette}, N.~D., {Naik}, S., \& {Patil}, M.~K. 2019, \mnras, 485, 1981

\bibitem[{{Vagshette} {$et~al$.}(2017){Vagshette}, {Naik}, {Patil}, \&
  {Sonkamble}}]{2017MNRAS.466.2054V}
{Vagshette}, N.~D., {Naik}, S., {Patil}, M.~K., \& {Sonkamble}, S.~S. 2017,
  \mnras, 466, 2054

\bibitem[{{Vagshette} {$et~al$.}(2012){Vagshette}, {Pandge}, {Pandey}, \&
  {Patil}}]{2012NewA...17..524V}
{Vagshette}, N.~D., {Pandge}, M.~B., {Pandey}, S.~K., \& {Patil}, M.~K. 2012,
  \na, 17, 524

\bibitem[{{Vagshette} {$et~al$.}(2016){Vagshette}, {Sonkamble}, {Naik}, \&
  {Patil}}]{2016MNRAS.461.1885V}
{Vagshette}, N.~D., {Sonkamble}, S.~S., {Naik}, S., \& {Patil}, M.~K. 2016,
  \mnras, 461, 1885

\bibitem[{{Wright}(2006)}]{2006PASP..118.1711W}
{Wright}, E.~L. 2006, \pasp, 118, 1711

\end{thebibliography}

\end{document}